\documentclass[sigconf]{nimeart}
\AtBeginDocument{%
  }

\setcopyright{cc}
\copyrightyear{2026}
\acmYear{2026}
\acmDOI{}

\acmConference[Goldsmiths Computing]{Technical Reports}{London, UK}

\acmISBN{}

\begin{document}

\title{Case Studies and Reflections on Agentic Software Engineering for Rapid Development of Digital Music Instruments}

\author{Matthew John Yee-King}

\affiliation{%
  \institution{Computing, Goldsmiths}
  \city{London}
  \country{UK}}
% \email{yeeking@gold}

\renewcommand{\shortauthors}{Trovato et al.}

\begin{abstract}
The article explores the use of agentic software engineering (ASE) in the development of innovative audio software. It begins with a review of background work that lays out the challenges of longevity, interoperability and barriers to entry in digital music instrument creation, explaining recent developments in ASE and highlighting the possibility that ASE can lower barriers to entry and facilitate creation of interoperable software with greater longevity. Following that, we present case studies wherein we used ASE technology in three distinct ways to develop audio software in the C++ language with the JUCE framework. In case study 1, we re-implement Laurie Spiegel's `Music Mouse' software as a native plugin. In case study 2, we translate Pachet's `Continuator' system from Python into a native plugin. In case study 3, we develop a new 3D user interface for an existing `tracker' sequencer using OpenGL. We describe the experiences of the human developer in the case studies via autoethnographic discussion of the prompt logs and snapshots of the software as it was developed. We identify effective practice for ASE use in this domain and suggest future steps for the work involving evaluation of the method with non-programmer musicians. 

\end{abstract}

\begin{teaserfigure}
  \includegraphics[width=\textwidth]{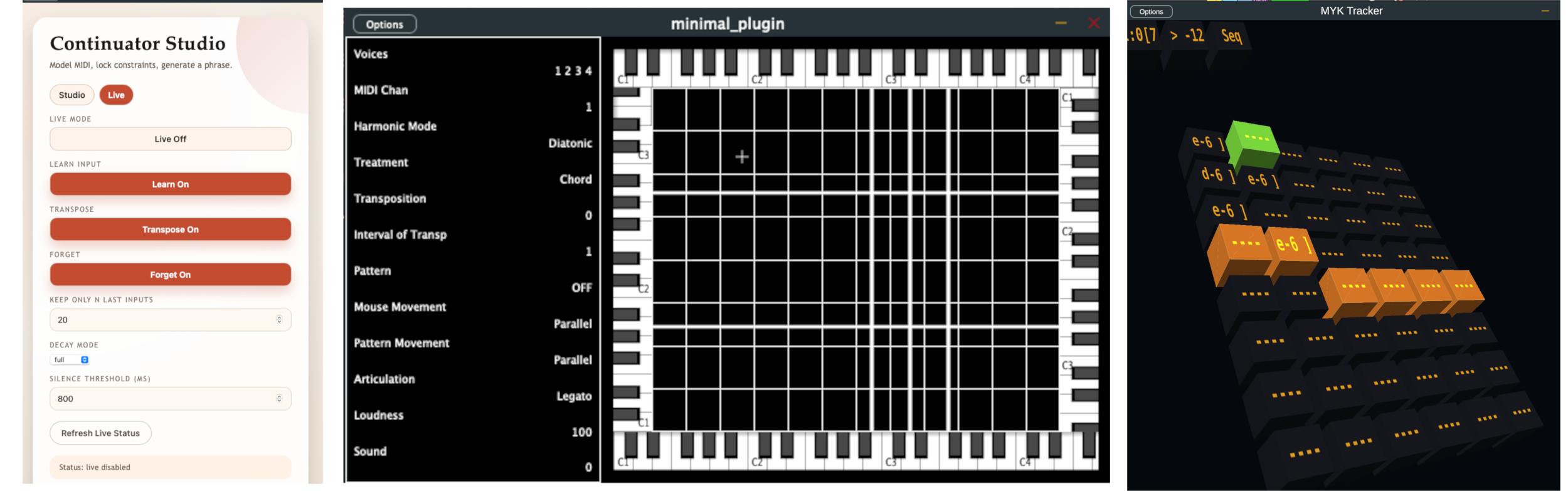}
  \caption{NIMEs programmed by Codex.}
  \Description{.}
  \label{fig:teaser}
\end{teaserfigure}

\maketitle

\section{Introduction}

In this paper, I explore the possibilities of agentic software engineering for audio software development. Agentic software engineering (ASE) involves humans developing software in collaboration with tool-using large language models which through an `agentic' layer, can develop and carry out plans. The human user inputs text, image and other media prompts and the agent enacts its plan by reading and analysing source code and documentation, creating and editing files, running build commands and so forth. The particular type of software I consider in the paper is `software digital music instruments' which I will henceforth refer to as SDMIs for brevity.

I focus on two areas of interest to the NIME community: the problem of longevity and interoperability and the challenge of lowering the barrier to entry for the potentially large community of non-programmers who wish to experiment with the creation of new music technology. I explore these questions through three distinct audio software development case studies: re-creating a well-known SDMI (Music Mouse) from the user manual and screenshots, translating an SDMI (Continuator) from Python to C++ and developing a new 3D OpenGL user interface for my own tracker sequencer. 

% What's the objective here?
%  - give them a bit of state of the art, tell them I know what I'm doing but acknowledge the issues perhaps? 
%  Maybe I need to clearly stake out the idea of a digital music instrument as software? 

This article was inspired by my recent experiences working with ASE tools to develop audio software. My experience has been that the tools very significantly accelerate audio software development cycles and that with an appropriate starting template and documentation, it is possible to develop interoperable audio software plugins in C++ entirely through natural language prompts, with limited knowledge of the underlying implementation. Thus, there is a lot of potential here for increasing interoperability of existing SDMIs (e.g. through translation into plugins) and lowering the barrier to entry to a wider community of SDMI creators. 

Whilst there is potential, there is limited published work exploring the use of ASE in the audio software domain, which is well-known to be a challenging area for software development. 
This is especially true when one considers the need to create interoperable, standardised software such as plugins with which  musicians are comfortable and familiar because plugins are  generally written in C++, a notoriously hard language to learn. To address this gap in the published work, I report on some initial experiments I have conducted to evaluate the capabilities of these emerging technologies. The contributions of the paper are as follows:

\begin{enumerate}
\item A description of a methodology for developing native audio software using ASE technology 
\item The application of the methodology to three distinct case studies and reflections on problems encountered and how they were overcome
\item A discussion of the potential of ASE technology to address well known problems in research-oriented SDMI development, namely longevity, interoperability and the barrier to entry for non-programmers
\end{enumerate}

\section{Previous work}
\subsection{Making NIME and SDMI development accessible}
Developing more accessible and efficient means to get from idea to prototype is not a new activity for computer music researchers. There is a long history of specialised programming languages and environments from Music-N through CSound, Max/MSP, PureData, SuperCollider and Faust and onto Bela, Livecoding languages and so forth, many of which are discussed by Dannenberg\cite{dannenbergLanguagesComputerMusic2018a}

Whilst Dannenberg claims that ``Computer music languages have enabled composers who are not software engineers to nevertheless use computers effectively'', anyone who has taught a class to composers about a textual language such as SuperCollider will be aware of the difficulty learners face in setting up the tools, then the brittleness and abstractness of the languages. For example, seasoned SuperCollider educator Collins discusses the challenges of transitioning learners from parameter editing to more complex program design\cite{collinsLiveCodingTeaching2016a}.

A lot of work has been done to address these issues: concerning the problems of brittleness and abstractness, visual languages such as Max/MSP are less brittle and less abstract and educators report significant success with non-programmers\cite{walzerAudioEducationTheory2020}. Concerning tool setup, a widely used solution is the use of `zero-setup', web-based environments such as the educationally focused MIMIC platform\cite{zbyszynskiWriteOnceRun2017,mccallumSupportingInteractiveMachine2020b}, the web-based version of the Tidal Cycles livecoding system Strudel\cite{mcleanMakingProgrammingLanguages2014a,roosStrudelLiveCoding2023,diapoulisTeachingStrudelYoung2024}, and the web interface provided with the Bela hardware platform\cite{donovanBuildingIDEEmbedded2017}.

Another problem associated with computer music languages, and a popular subject in the literature, is that of longevity and sustainability of the software created using them. For example, in a review of 40 years of NIMEs, Sullivan and Wanderley stated that ``a DMI may not be viable unless basic stability, reliability and compatibility standards have been met''\cite{sullivanStabilityReliabilityCompatibility2018}. An emerging trend which addresses compatibility is the development of plugins which interoperate with standard audio software used by musicians\cite{yee-kingStrategiesBuildingAIenhanced2024}. Examples are the DrumGAN and RAVE VSTs\cite{nistalDrumGANVSTPlugin2022,caillonStreamableNeuralAudio2022}.

The problem with plugins is that they are typically written in the C++ language which is notoriously difficult. To quote  Bjarne Stroustrup, the creator of C++: ``C makes it easy to shoot yourself in the foot; C++ makes it harder, but when you do it blows your whole leg off''\footnote{https://www.stroustrup.com/quotes.html}. Given this, it seems reasonable to assert that the barrier to entry for non-programmer musicians to plugin development is higher than that for, say, Max/MSP. 

Though we should note that plugins are not the only route to interoperability, for example, `Max for Live' allows Max/MSP patches to run inside Ableton Live\footnote{https://cycling74.com/products/maxforlive}; the Faust IDE allows users to develop in the Faust language and to use the web platform to directly compile their program into a binary VST plugin (as well as many other formats)\cite{letzWHATSNEWFAUST2024}. But if musicians wish to develop software that integrates with their other music tools, C++, Max for Live and Faust all share the requirement to operate and express ideas in the unfamiliar medium of a programming language (even if it is a visual language like Max). 

\subsection{From code completion to agentic software engineering}

Vibe-coding became the Collins English Dictionary word of the year in 2025, which defined it as ``the use of artificial intelligence prompted by natural language to assist with the writing of computer code''\cite{collinsenglishdictionaryVIBECODINGDefinition2025}. 
LLM-supported programming did not start in 2025 - OpenAI described the first Codex model in 2021 as a ``a GPT language model fine-tuned on publicly available code from GitHub''. 
% Codex was also the model behind the then-nascent copilot system\cite{chenEvaluatingLargeLanguage2021a}.
%
In the HumanEval benchmark, this initial version of Codex was able to provide functionally correct Python code in response to docstrings (i.e. to write code from natural language descriptions) for 28.8\% of the problem set. As of 2025, GPT-5 (version unspecified) scores 93.5\%, with the Kimi-K2 open-weights model surpassing it at 94.5\% \footnote{https://llm-stats.com/benchmarks/humaneval, https://huggingface.co/moonshotai/Kimi-K2-Instruct-0905}.

Through the 2020s, the capabilities of LLM-powered coding tools rapidly expanded with the addition of features such as codebase indexing and command-calling with custom-designed IDEs such as Cursor\footnote{https://cursor.com/}. These extended capabilities of LLMs whereby they can `do things' in a computer system are referred to as tool-use\cite{qinToolLLMFacilitatingLarge2023}. 
There are different ways to present a set of tools to an LLM; for example, Anthropic defined the Model Context Protocol (MCP) in 2024 and it allows the description of tools and their execution\footnote{https://www.anthropic.com/news/model-context-protocol}\cite{ntousakisSecuringMCPbasedAgent2025}. 
A further innovation is the reasoning LLM, which carries out an iterated process of reasoning prior to emitting a result\cite{plaatMultiStepReasoningLarge2026}. Finally, `Agentic AI' combines LLMs, tool-use and reasoning into a software agent which can adaptively develop and carry out plans on a computer system\cite{aboualiAgenticAIComprehensive2025}
%MCP is a standard protocol via which you can expose tools and resources to language models, whereby you install MCP services for different tools in your environment and the language model can then query and use those tools. For example, an MCP server could provide the capability to search Wikipedia or to call commands in a terminal. There is a list of MCP servers here\footnote{https://github.com/punkpeye/awesome-mcp-servers} and there are MCP servers that let agents interact with Max/MSP\footnote{https://github.com/tiianhk/MaxMSP-MCP-Server} and Reaper\footnote{https://github.com/yeeking/reaper-mcp-server}. 
% Building on tool-use, Chain-of-Thought or reasoning models are able to carry out a reasoning process prior to emitting the result\cite{plaatMultiStepReasoningLarge2026}. The most recent development is referred to as agentic AI, wherein traditional software agents, defined as ``autonomous software entities engineered for goal-directed task execution within bounded digital environments''\cite{sapkotaAIAgentsVs2026} are combined with tool-using, reasoning LLMs to allow for highly adaptive plan creation and enactment. 

% Most recently, wrapping the tool-using, reasoning language model with an `agentic' layer allows the creation and adaptive enactment of plans based around the use of a sequence of tools. 

%
For programming tasks, the current manifestation of all these capabilities is referred to as agentic software engineering\cite{hassanAgenticSoftwareEngineering2025}. In October 2025, OpenAI made their latest iteration of Codex `generally available' (i.e. to lower-tier, paying customers), and it supports multi-modal agentic software development using build tools on your local machine\footnote{https://openai.com/index/introducing-gpt-5-2-codex/}. The Codex agent itself is open source\footnote{https://github.com/openai/codex} but it calls out to dynamically selected, closed source, closed weights language models running in the cloud. Codex is the tool used in the work presented in this article, but there are other ASE systems available such as Anthropic's Claude Code\footnote{https://github.com/anthropics/claude-code}. Tt is even possible to run the agent and the models on your own machine with tools such as Roocode\footnote{https://github.com/RooCodeInc/Roo-Code}, LMStudio\footnote{https://github.com/lmstudio-ai}, and open weights models\footnote{https://huggingface.co/}, though you will need an expensive system. 

%
%

% Codex and Claude Code both execute partially on your local machine and partially in the cloud but it is also possible to run the complete agentic software engineering workflow on your own hardware. You will need a powerful computer, a coding agent with integration to your IDE e.g. Roocode\footnote{https://github.com/RooCodeInc/Roo-Code}, a tools-capable language model host e.g. LMStudio\footnote{https://github.com/lmstudio-ai}, possibly some MCP servers and some open weights models\footnote{https://huggingface.co/}. 

%
Compared to the chat-style interaction and code completion available in the first generation of `vibe-coding' systems, the full agentic software engineering workflow is remarkably powerful and, perhaps most importantly for the widening access theme in this paper, highly autonomous. The author's experiments developing audio software in C++ with Codex in late 2025 inspired this paper. 

\subsection{Using language models to write audio software}
There is an emerging body of work discussing ASE in general, but there is limited work discussing the use of LLM technology for audio software development. Zhane et al. benchmarked LLMs in their abilities to work in visual dataflow languages, generating programs in Max/MSP, MaxPy, JavaScript/ Web Audio API and their own visual language Wavir\cite{zhangBenchmarkingLLMCode2024}. 
% They used the LLM to carry out both ``highly defined, low-level tasks'' and ``open-ended creative code possibilities'' and used human evaluators to measure correctness and complexity of the solutions. 
%
The work presented here differs in that we discuss the creation of large, complex programs in C++ through multi-stage agentic methods, instead of just working on a prompt-to-code basis. 
Smith discussed the use of LLMs to convert the filter design Python code in scipy into a C++ library with the objective of creating an audio plugin that allows you to design your own filters\cite{iiiUsingAIPort2024}. 
%He developed a workflow that firstly generates Python code, then translates to C++, on the basis that LLMs are trained on a lot of Python and it is a simple language, and that translation from a working program would be more effective than going straight from prompt to C++. 
% He noted that in 2024, chatbots ``are not yet good at signal processing (even simple polynomial algebra), and tend to fall apart on low-level signal-processing details'', but ``Translating Python to C++ for real-time use is greatly facilitated by Chatbots''. 
Similarly, one of the case studies reported below explores translation from Python to C++.

\section{Method: three case studies in agentic audio software engineering}

\begin{figure}[h]
  \centering
  \includegraphics[width=\linewidth]{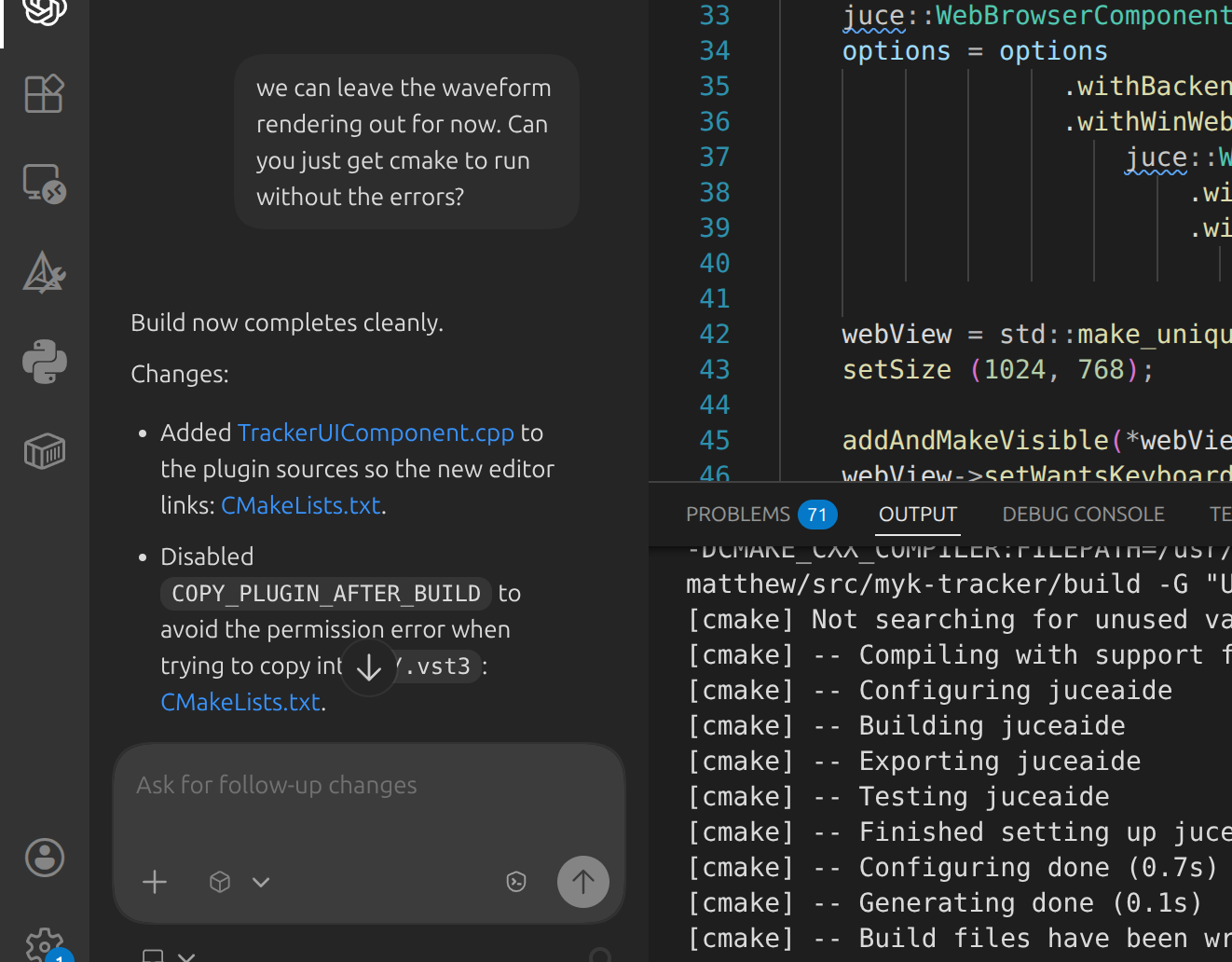}
  \caption{Interacting with codex in the VSCode extension.}
  \Description{The user interface of the codex extension in VSCode.}
  \label{img:codex}
\end{figure}

\begin{table}[]
\centering
\caption{Three case studies using agentic software engineering to develop audio software}
\label{tab:case-studies}
\resizebox{\columnwidth}{!}{%
\begin{tabular}{|l|l|l|l|}
\hline
\textbf{Case study}        & \textbf{Process type} & \textbf{Template code}      & \textbf{Documentation input} \\ \hline
Music Mouse                & From manual and images          & JUCE CMake project          & Manual, UI screenshots       \\ \hline
Continuator                & Translate             & JUCE CMake/ HTML UI project & Python code                  \\ \hline
MYK-Tracker & New UI                & JUCE CMake project          & C++ code                     \\ \hline
\end{tabular}%
}
\end{table}

The basic process I carried out in this research was to identify three interesting software projects or case studies, then to use ASE technology to interactively develop the software, whilst recording data about my experience. More details of the data capture are presented below. Each case study explores a distinct `challenge', as described in table \ref{tab:case-studies}. The challenges are 1) create a plugin from scratch to a specification, 2) translate from a Python web application to a C++ plugin and 3) create a new 3D OpenGL interface for an existing audio program. 

In acknowledgement of the `longevity of NIMEs' theme of this paper, I decided that two of the case studies would involve the re-implementation of well-known computer music systems from the past. This has been a theme in previous NIME papers such as Masu et al. who re-implemented NIMEs from past conferences in 2023\cite{masuNIMEReflectingImportance31}, and Calegario et al. who evaluated documentation for a range of NIMEs to evaluate reproducibility\cite{calegarioDocumentationReplicabilityNIME2021}.

% \subsection{Method}

\begin{figure}[h]
  \centering
  \includegraphics[width=\linewidth]{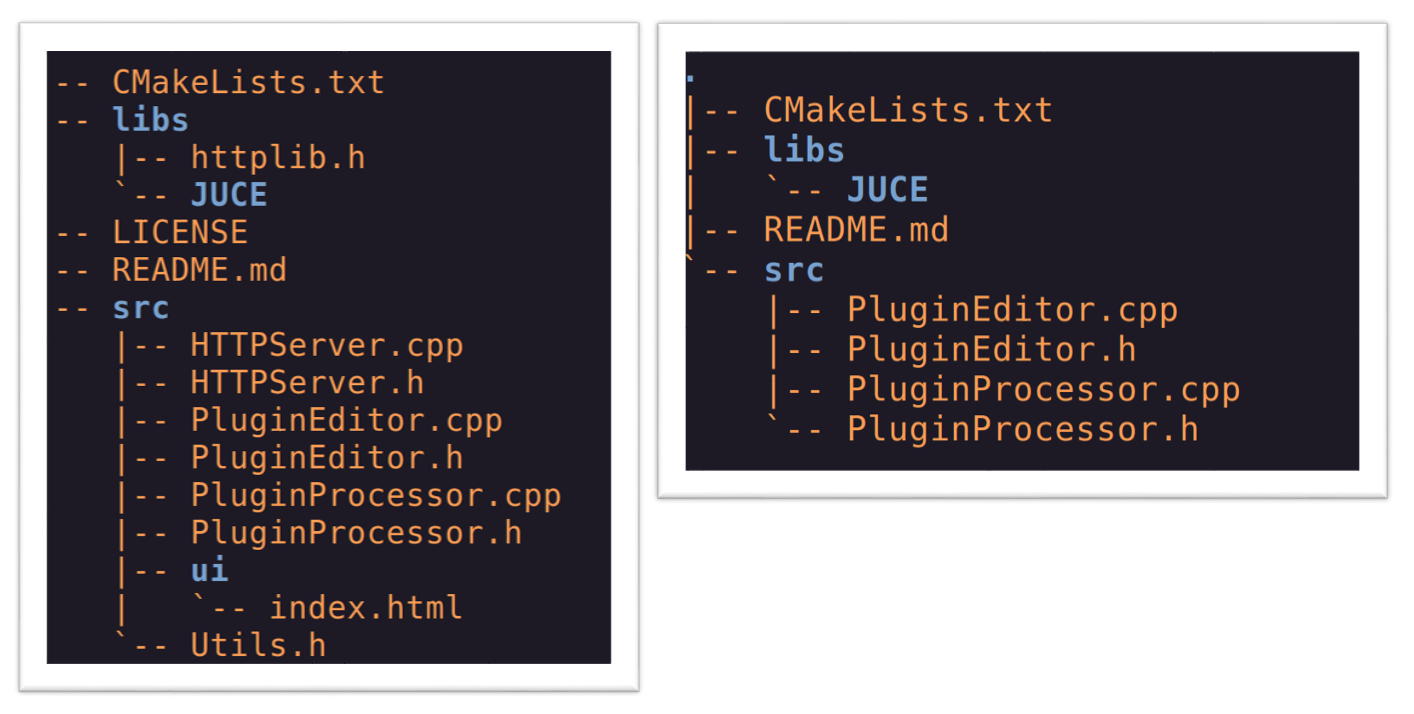}
  \includegraphics[width=\linewidth]{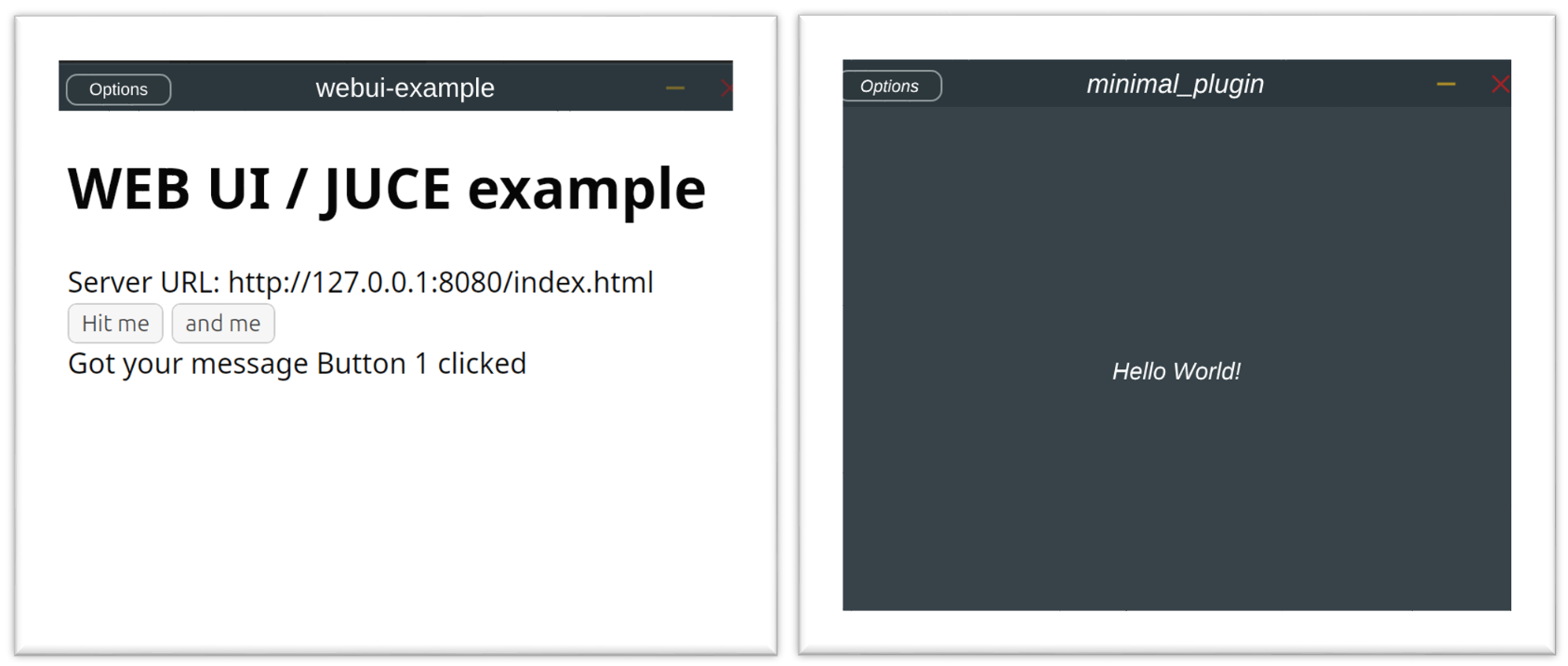}
  \caption{Project file structure for WebView starter template (top left) and JUCE UI template (tio right). User interfaces for WebView (bottom left) and JUCE UI (bottom right). The JUCE git repository sits in the libs folder}
  \Description{Two user interfaces and project structure for different starter templates.}
  \label{img:templates}
\end{figure}

\subsection{Programming environment and project templates}
I used the Codex ASE system to develop the software described in the case studies. You can see a screenshot of the Codex extension running inside VSCode in figure \ref{img:codex}. I worked in the C++ language on top of the JUCE framework\footnote{https://juce.com/}. JUCE is a widely used, open-source, cross-platform framework which provides project build management and a C++ class library for audio and GUI development. JUCE allows the development of VST, AudioUnit and other plugin formats as well as standalone and command-line audio programs. 

From previous work[anon], I had two template projects which I used as starting points for development. The most basic template was written completely in C++ with the user interface built using the JUCE GUI classes. The other template used an HTML/CSS user interface which would be embedded to the native interface via a WebView component. The web UI interacts with the audio code via REST API calls and JavaScript callbacks. This workflow is an emerging trend in plugin development which allows developers to develop their interfaces in HTML and CSS instead of the more limited and obscure native GUI libraries available. I expected Codex to be able to work very well in HTML and CSS, as it would have featured much more strongly in its training data than, for example, the JUCE UI library. Figure \ref{img:templates} shows the file structure of the two templates and the initial user interfaces they provided. The projects included a CMake build configuration that would allow them to be built on the command line. They also included a complete copy of the JUCE framework in a `libs' sub-folder with examples, documentation and library code that Codex could explore. Prompting of Codex was done with text and images.

% The projects consisted of a blank audio plugin or standalone application-type program with a user interface class (PluginEditor) and an audio processor class (PluginProcessor). The projects would always have a copy of the complete JUCE GitHub repository in a sub-folder which the agent could read and analyse. 

%

%
% The projects were underpinned with the CMake tool which enables the complete building and testing workflow to be carried out via the command-line. It also allows for a unified cross-platform workflow. Since the coding agent can run commands and analyse their output, using CMake was more straightforward than having the agent interact with an IDE's build controls which might not be fully accessibly from a command line perspective. 

% During the process of development, I used git to log progress. Each major phase of development was carried out on a branch, and if successful, merged into the main branch. If an attempt failed, the branch could be abandoned. 
% Sometimes, I would use a standard chat interaction outside of the ASE environment to work up a prompt for the agent. This allowed me to interactively develop a clear specification without polluting the context for the interaction with the agent. It also allowed me to use the more expanded capabilities of the web-based chat such as file upload, web search and image generation to develop the ASE prompt. 

\subsection{Data capture and analysis}

I aimed to use a light form of auto-ethnography to analyse my experiences and thoughts when developing the software. To this end, I collected the following data whilst carrying out the development:

\begin{enumerate}
  \item Screencast recordings of my computer screen
  \item Audio recordings with transcripts from a mic that allowed me to comment on what I was doing and experiencing whilst working with the agent 
  \item Detailed, timestamped logs of all user and agent actions in JSON format, including the prompt sequence 
  % \item The git logs which show the time spent, branches and stages of development
  \item The source code itself 
\end{enumerate}

To analyse the data, I planned to watch the screencasts, to listen to the commentary, to read the generated source code and to use automated tools to analyse and visualise the log data. 

% watched the screencasts. I gathered screenshots of the user interfaces of the programs as they developed. I wrote notes that identified major phases of the development, along with timings and prompts. 

% I planned to analyse the audio transcripts to identify themes and phases in my development experience. 

\subsection{Case study 1: re-creation of a classic SDMI from the user manual and screenshots (Music Mouse)}
\begin{figure}[h]
  \centering
  \includegraphics[width=\linewidth]{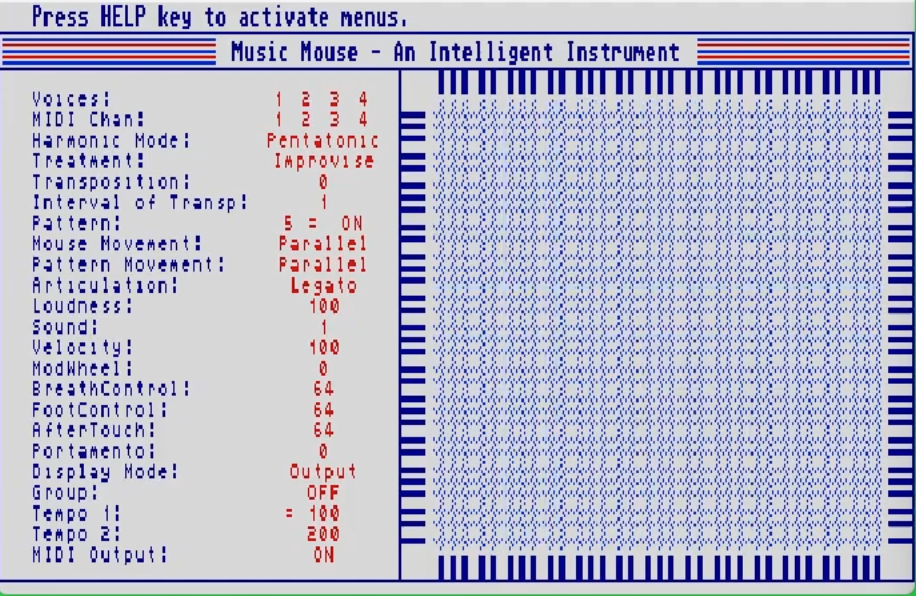}
  \caption{Music Mouse running on an emulated Atari ST; Music Mouse was also available for Amiga and MacOS.}
  \Description{The user interface of the Music Mouse software.}
\end{figure}

The first case study involved re-implementing an SDMI from scratch using the basic JUCE UI plugin template. I selected an SDMI from the literature that was a) well documented, b) software only, not hardware, and c) had an interesting/ distinctive user interface. I selected Laurie Spiegel's Music Mouse software for re-implementation as it meets the three criteria\cite{spiegelMusicMouseIntelligent1986}. In Music Mouse, the user moves the mouse around in a note grid and the system generates chord and single note events according to a range of parameter settings. 
%There is also a sequencer mode wherein the notes are generated in rhythmic patterns. 
Music Mouse was commercially available during the 80s and 90s for the Atari ST, Commodore Amiga and MacOS platforms. Whilst the original last release, V2.1.2 Classic was made available in 2005, I discovered during the preparation of this paper that Eventide planned to re-release Music Mouse in 2026\footnote{https://www.eventideaudio.com/software/music-mouse/} - mit dem Zeitgeist leben!

\subsection{Case study 2: Python to C++ native translation of classic SDMI (Continuator) }

Pachet's Continuator is a Markov-based system designed for interactive musical sequence generation, originally created in the early 2000s\cite{pachetContinuatorMusicalInteraction2003}.
Later developments involved advanced techniques for steering sequence generation using constraints\cite{pachetMarkovConstraintsSteerable2011}. 
Continuator was implemented in the Java language, with no source code available, but in 2025 Pachet re-implemented the system in Python with a gradio web interface and released the code on GitHub\footnote{https://github.com/fpachet/continuator}. In this case study, I converted the Python code into a C++ VST/AudioUnit plugin using the WebView template, since the Python version also had a web interface. 

\subsection{Case study 3: user interface conversion from JUCE to OpenGL ([anon]-tracker)}

In the final case study, I investigated the capabilities of ASE techniques in converting the user interface for a fairly complex `tracker' sequencer from JUCE UI components to an OpenGL UI. This would allow me to observe the agent working in a pre-exiting, moderately complex codebase. Trackers are a form of music  sequencer often associated with the Chiptune scene, but they were also used extensively in 1990s UK electronic music by artists such as Black Dog\cite{reunanenTrackersRiseBloom}. I created the open-source tracker software some time ago with purely human C++ programming, aiming to run it on a small system such as a Raspberry Pi\footnote{anon}. In the original project, I started out with a command-line curses interface without JUCE and later adapted that to a JUCE plugin with JUCE UI components so I could more easily implement DSP functionality. In the case study, I planned to work with ASE to implement a completely new user interface. 
%Figure \ref{img:} shows the somewhat un-inspiring Curses and JUCE UIs - I hoped that ASE would unlock unfamiliar low-level OpenGL programming capabilties and provide a more engaging user interface. 

\section{Results and analysis}
\begin{figure*}[h]
  \centering

  \includegraphics[width=\linewidth]{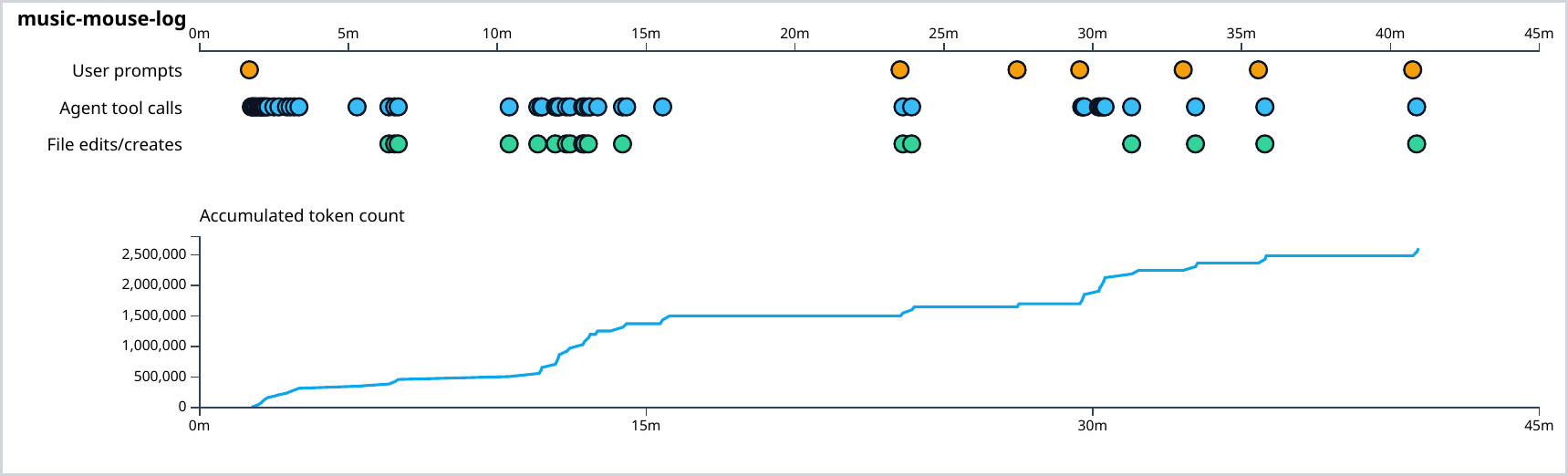}
  \includegraphics[width=\linewidth]{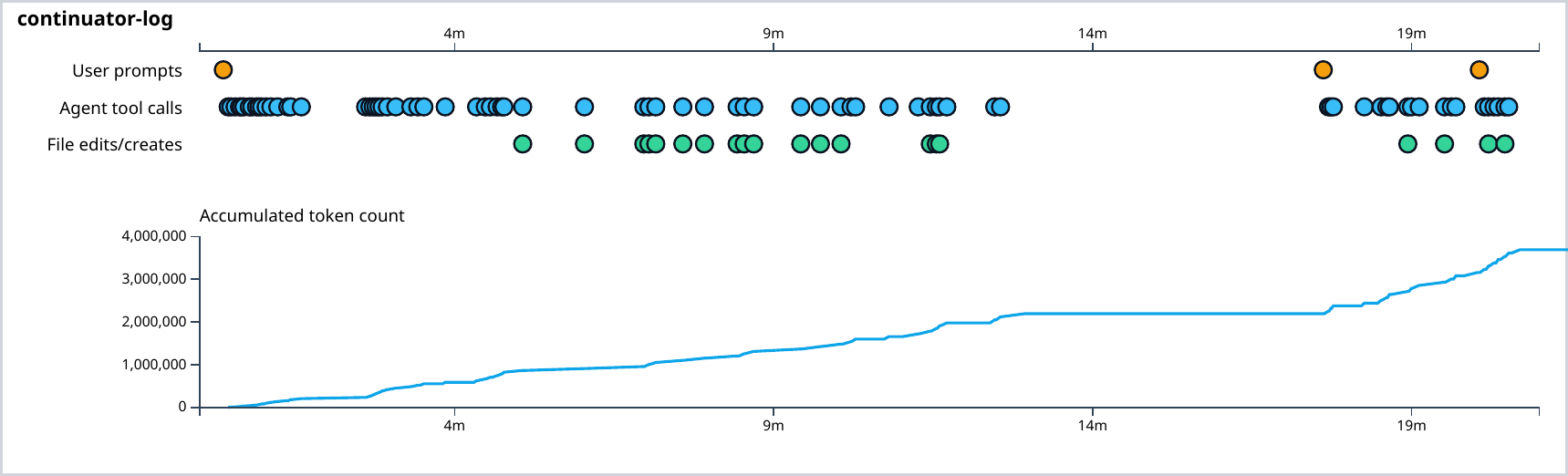}
  \includegraphics[width=\linewidth]{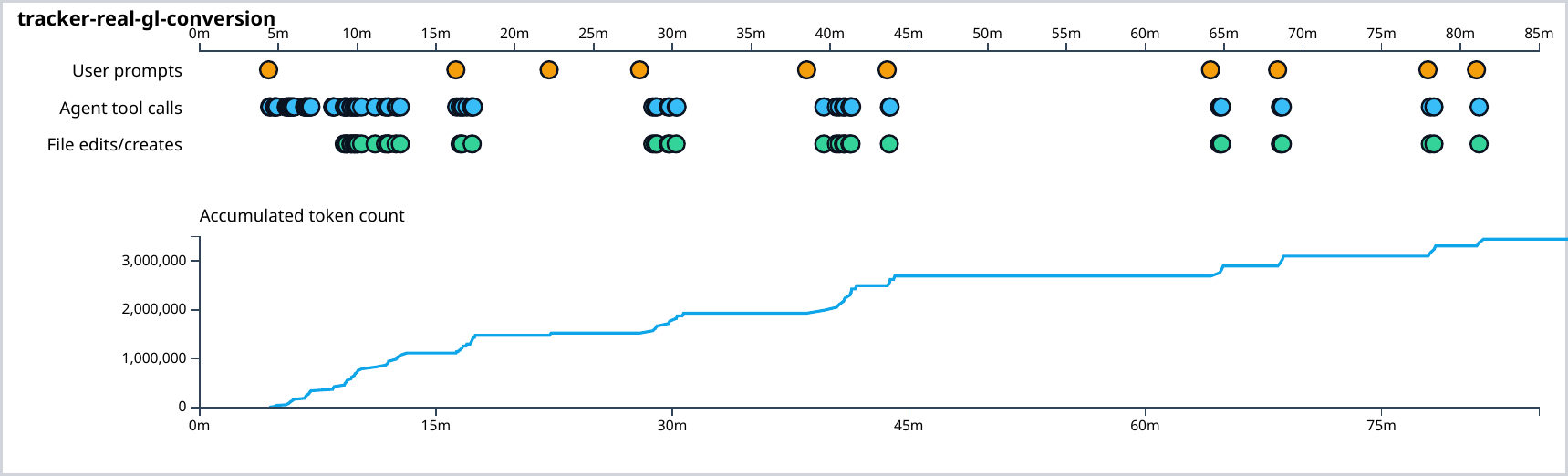}

  \caption{Timelines for the three case studies: Music Mouse (top), Continuator (middle), OpenGL tracker interface (bottom). Rows of circles show actions: first row prompts, second row tool calls, third row file edit actions. Graph shows accumulated tokens sent to the LLM. Codex was used to prepare this visualisation.}
  \Description{Timelines of agentic software engineering sessions.}
  \label{img:timelines}
\end{figure*}

In figure \ref{img:timelines}, I present visualisations of the activities that took place in each case study. There you can see when I wrote prompts, when Codex used tools and the number of tokens processed over time. In the following subsections, I will describe my activity and observations during the three case studies. 

% Full prompt logs are included in the appendix, except for the images sent. 

\subsection{Case study 1: Music Mouse}
\begin{figure}[h]
  \centering
  \includegraphics[width=0.45\linewidth]{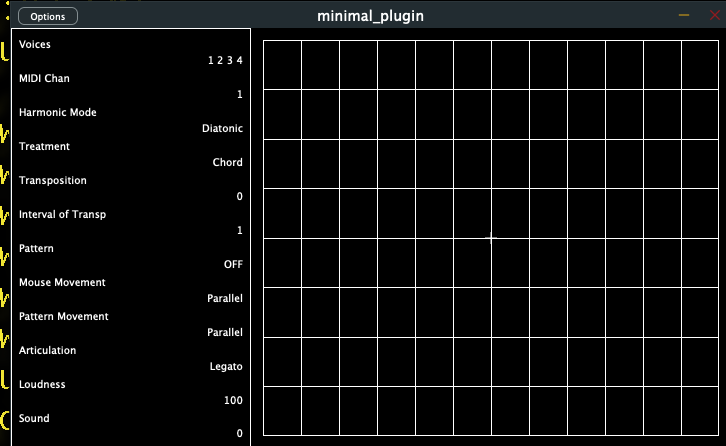}
  \includegraphics[width=0.45\linewidth]{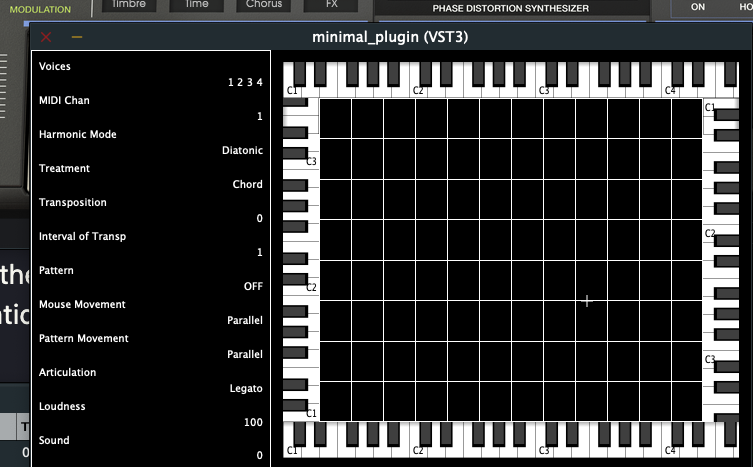}
  \includegraphics[width=0.45\linewidth]{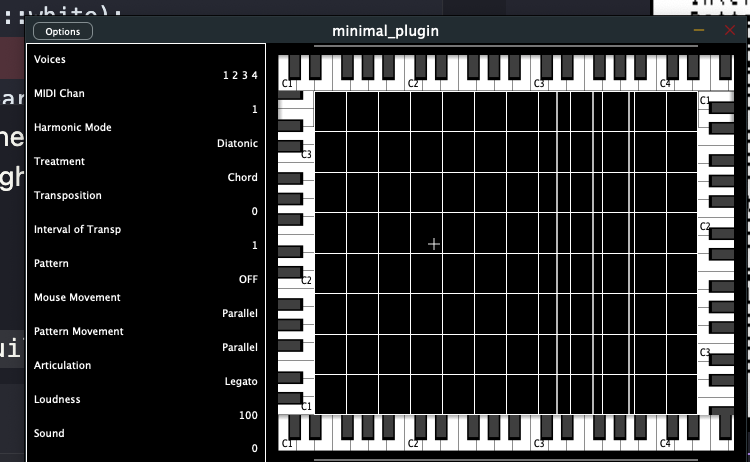}
  \includegraphics[width=0.45\linewidth]{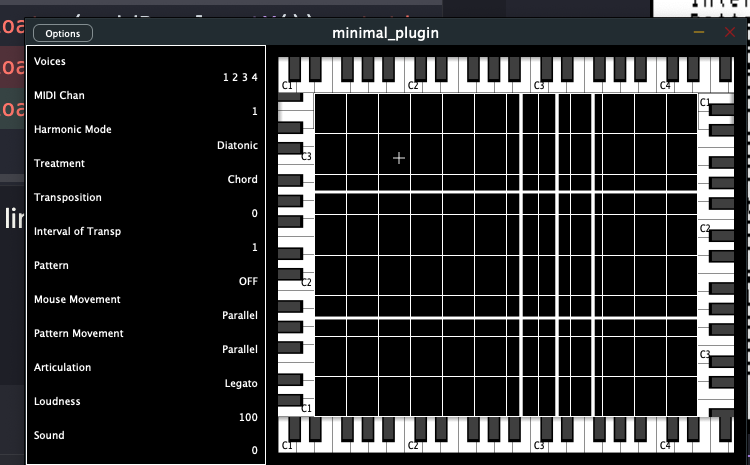}
  \caption{Four iterations of the Music Mouse re-implementation, clockwise from top left. V1 approximates the user interface with a mostly working mouse to MIDI interaction and keyboard controls; V2 includes the piano display, V3 adds graphical indications of the note positions and V4 enhances the horizontal and vertical bars. }
  \Description{The user interface of AI re-implementation of the Music Mouse software.}
  \label{img:mm-codex}
\end{figure}

The Music Mouse session is shown at the top of figure \ref{img:timelines}. The visualisation begins with the first prompt, but prior to that I spent around 20 minutes  preparing the project repository. I located a copy of the Music Mouse manual in HTML format on the Wayback Machine which I downloaded and placed into a `docs' folder in the project folder. Next, I searched the Internet for images of the original Music Mouse software, as there were no screenshots in the manual. I found an image of the MacOS version of the program and an image representing an Atari ST keyboard shortcut `cheat-sheet'. I placed these in the docs folder as well. Next, I prepared the example template, which was the basic JUCE / CMake plugin project consisting of a CMakeLists.txt file plus the cpp and header files for the user interface and audio modules. I cloned the JUCE repository into a `libs' sub folder and ran a test build to verify everything worked. 

I was now ready to prepare the first prompt. In the prompt I explained the objective, to re-implement Music Mouse, and described the structure of the repository including the location of the user manual and the JUCE repository. I also highlighted the importance of the mouse interaction and the keyboard controls. I included the MacOS screenshot with the prompt. 

About eight minutes later Codex had designed and carried out its first plan, which involved parsing the manual using Python commands, implementing the mouse and keyboard interaction models, and building a MIDI event queueing system. In the next phase I tested the plugin by instantiating it in a plugin hosting environment. I wired it into a MIDI logger plugin to show the MIDI events it was generating, and an emulated Casio CZ100 synthesizer (as mentioned in the Music Mouse documentation I had read) so I could hear the notes. I tried various keyboard shortcuts and mouse movements and was able to control transposition and chord generation. 

Then I developed several further prompts to improve firstly the note sending behaviour and then the way the user interface indicated the notes being played. I began to use an efficient workflow wherein I would be testing the plugin whilst Codex was working on the next build, then writing a new prompt ready to set off the next development cycle. Overall, the session lasted around 50 minutes and resulted in what seemed to be a reasonable implementation of the core keyboard and mouse interactions for Music Mouse. Main phases in the session were: project preparation, major prompt, plugin testing then smaller prompting and testing iterations.  

% I do not plan to release the code as the program is copyrighted and, as noted earlier, it is concidentally being re-released in 2026 as a commercial product by Eventide. Figure \ref{img:mm-codex} shows four iterations of the re-implementation of the Music Mouse program captured as the ASE process was carried out. 

\subsection{Case study 2: Continuator}

\begin{figure}[h]
  \centering
  \includegraphics[width=0.45\linewidth]{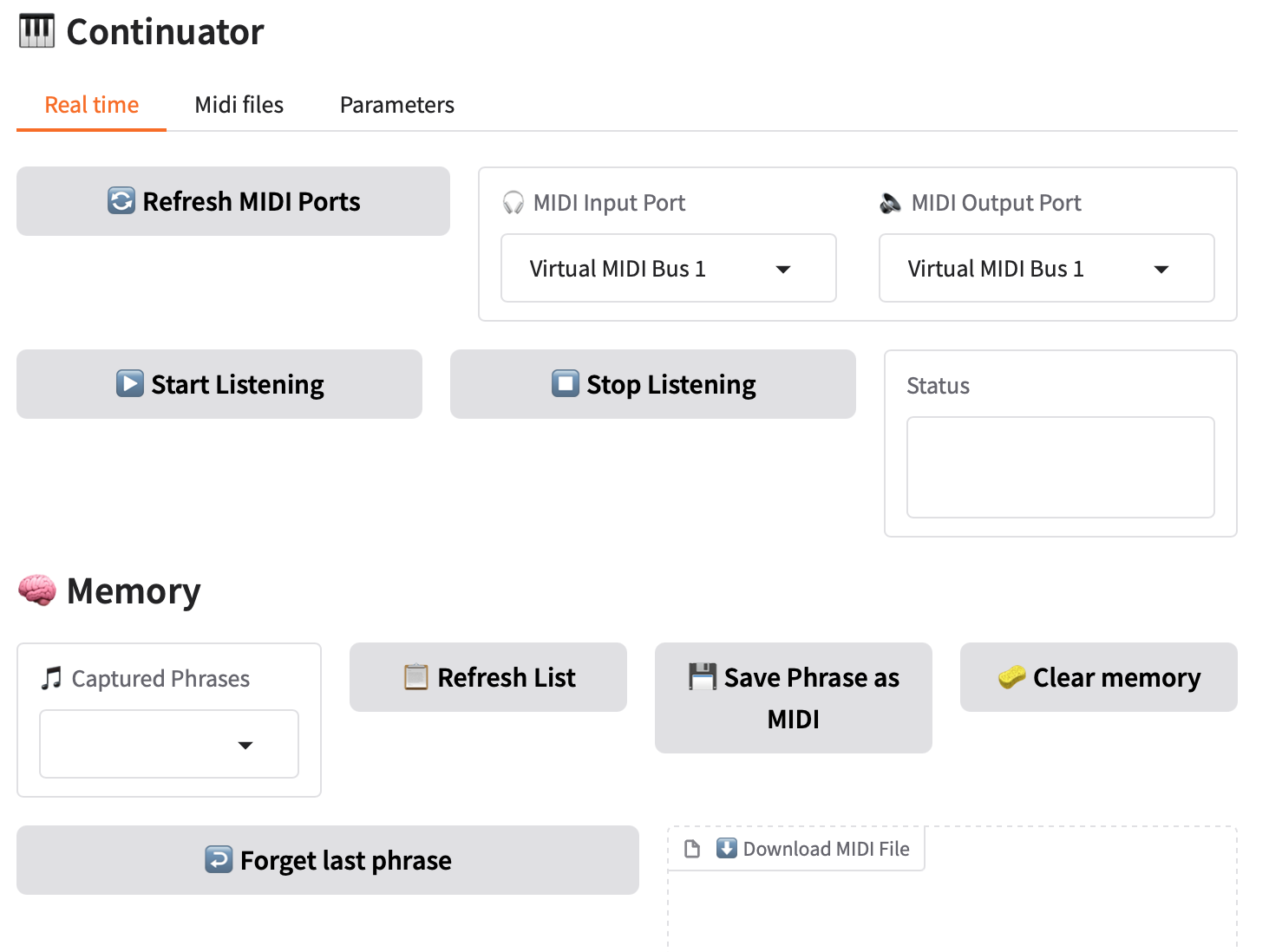}
  \includegraphics[width=0.45\linewidth]{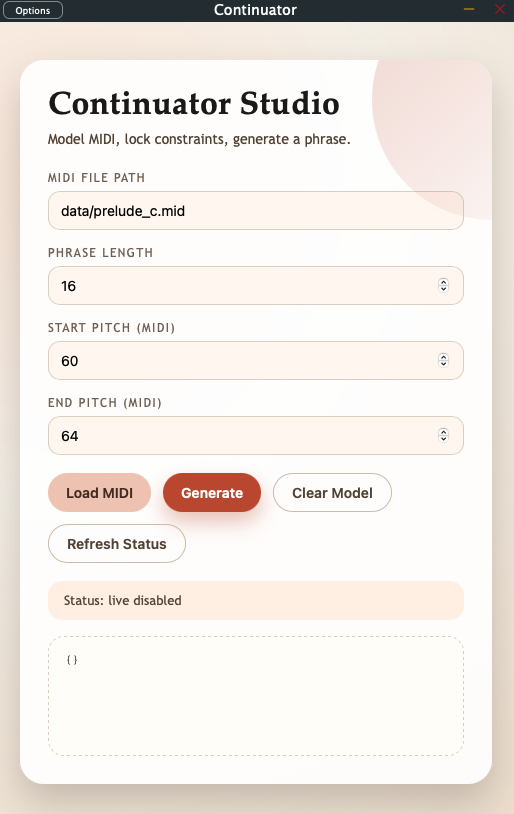}
  \includegraphics[width=0.45\linewidth]{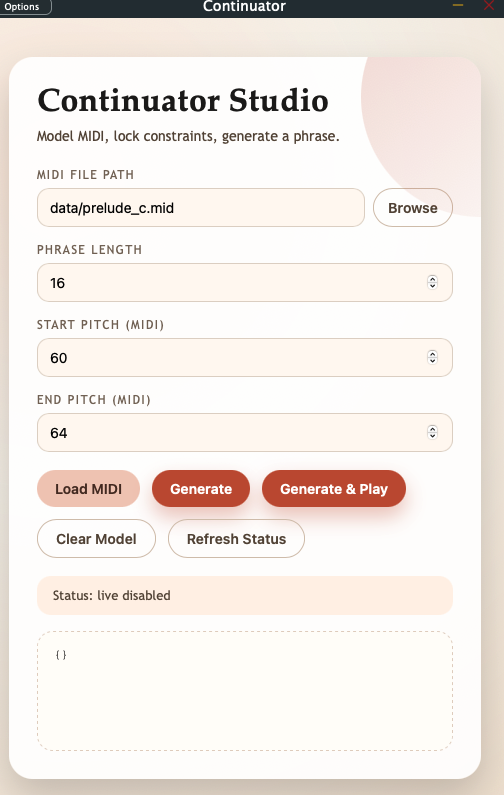}
  \includegraphics[width=0.45\linewidth]{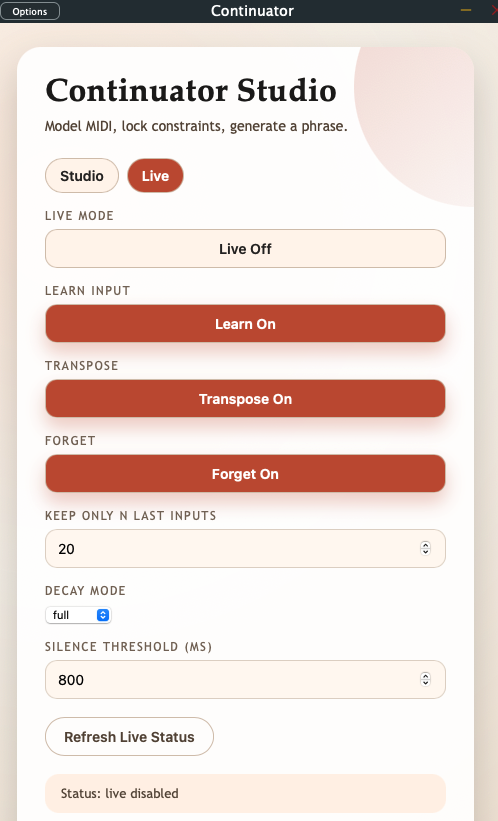}
  \caption{Original Python Continuator interface (left) then three iterations of the Codex written version from left to right. V1 can model MIDI files and generate constrained output, V2 can generate the output as correctly timed MIDI messages and V3 implements the full Continuator call and response interaction. }
  \Description{.}
  \label{img:cont}
\end{figure}

Figure \ref{img:cont} illustrates the progression of the Continuator plugin user interface during the session, including the original Python web interface. As for Music Mouse, the first phase involved setting up the project folder, except now I used the WebView plugin template. I added a clone of the Python Continuator repository and an example of a command-line JUCE unit-testing program. The unit-testing example code was there because I planned to include a request for unit-tests of the Continuator algorithm in my prompts. 
Next, I developed and sent the first prompt which explained briefly what Continuator does and that I wanted a translation of the Python project into a plugin with a web interface. I also included a request to write unit tests to validate the algorithm. After 12 minutes of work, Codex returned; I ran the plugin and the core of the algorithm appeared to be in place in that you could load and model a MIDI file and then generate note events from the model in text format. 

I re-organised the files in the repository to my liking and asked Codex to fix the build with the new arrangements. I returned to the project some days later for the next phase wherein my aim was to get the plugin to not just generate its output in text format but also to emit correctly timed MIDI messages. One prompt and four minutes later, the plugin was emitting MIDI messages. I tested in the Reaper DAW, connecting the output to a Rhodes electric piano sound. The second stage of the session is not shown in figure \ref{img:timelines} but it follows a similar pattern to the other timelines, with the first prompt taking a longer time to process followed by multiple, smaller prompt processing stages. 

At this point, I felt compelled to contact Continuator's original creator Francois Pachet (with whom I occasionally correspond), so I sent him a video of the plugin in action. He was very impressed to find out that it wasn't just a front-end for the Python code but a complete, native translation. I worked with him to get the CMake build running on his system which was very straightforward and he easily ran the AudioUnit version of the plugin in the Logic Pro DAW. He highlighted some interesting problems such as what would be the canonical version to which bug fixes were applied - the Python or C++ implementation? Also how do you know it is functioning correctly? We discussed next steps and he was keen to see the complete Continuator `call and response' behaviour implemented so I worked with Codex to implement that which took two further phases and a few minutes. I ended the session there. Compared to Music Mouse, the Continuator case study took place over several sessions, but it followed a similar pattern of preparation, major prompt, prompt processing, testing then multiple smaller prompt-test iterations. 

\subsection{Case study 3: OpenGL tracker interface}

\begin{figure}[h]
  \centering
  \includegraphics[width=0.45\linewidth]{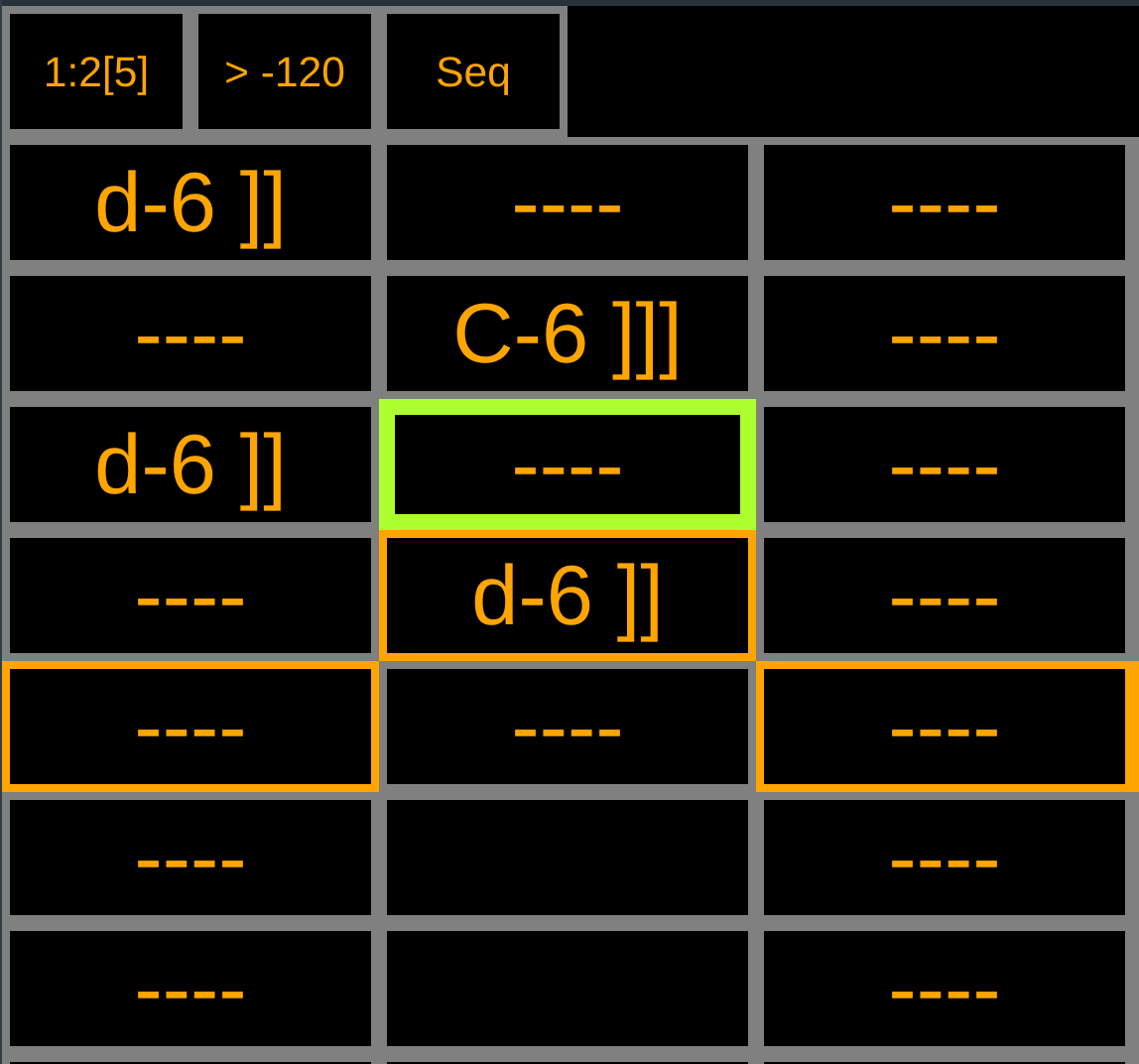}
  \includegraphics[width=0.45\linewidth]{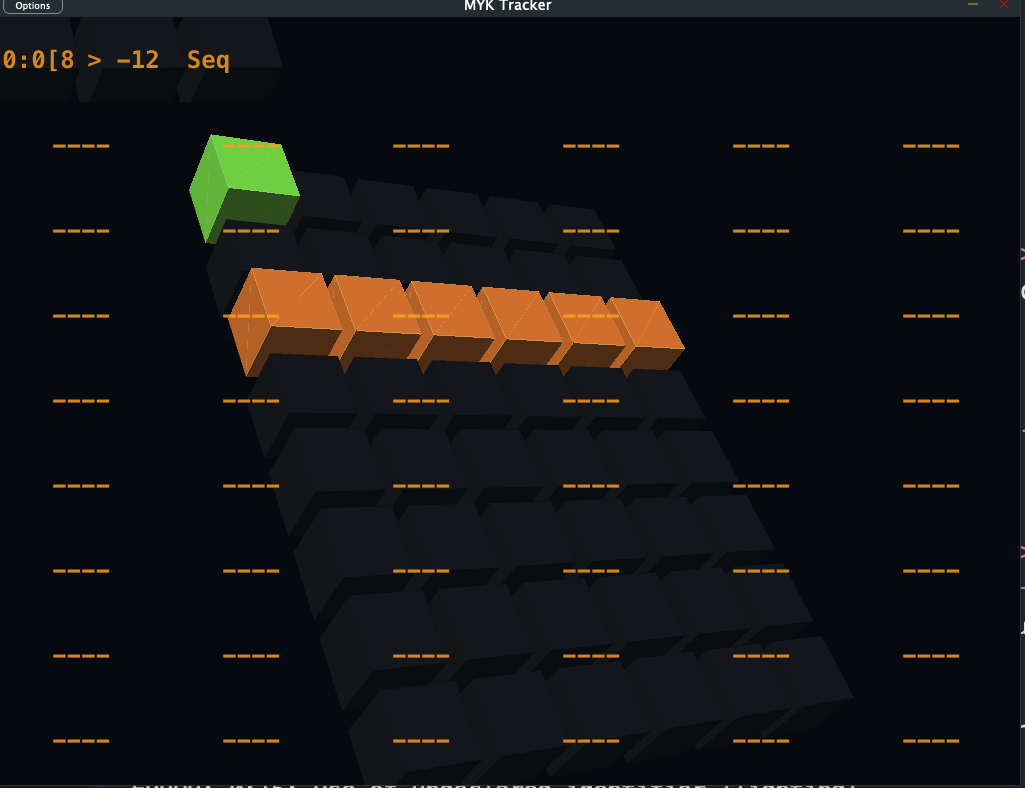}
  \includegraphics[width=0.45\linewidth]{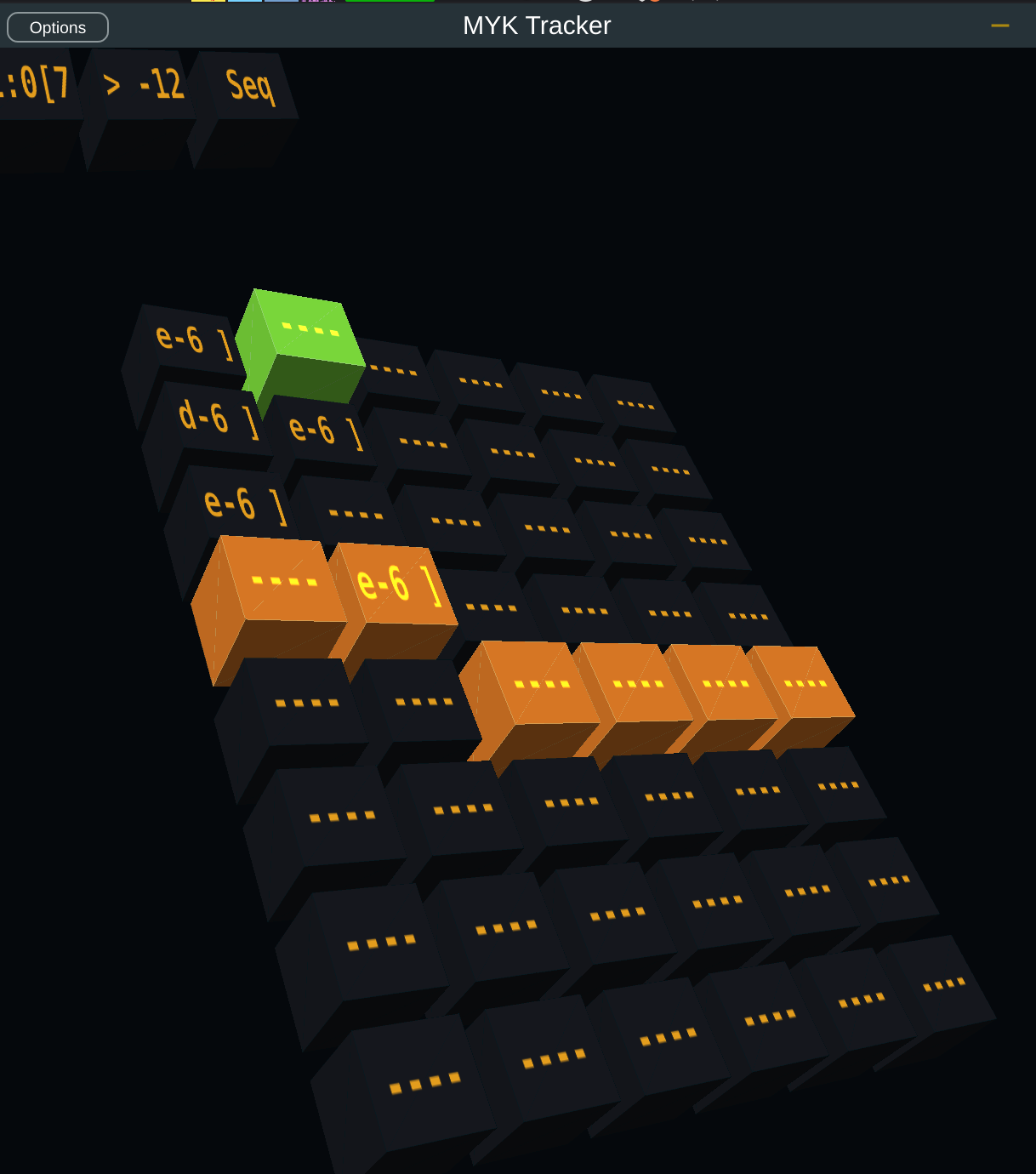}
  \includegraphics[width=0.45\linewidth]{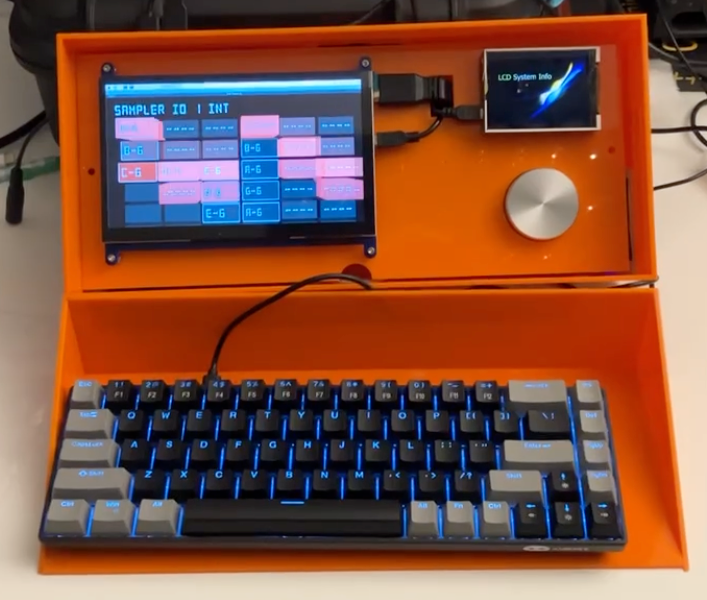}

  \caption{Four iterations of the tracker UI: original JUCE version (far left), V1 with text overlay problem (centre left), V2 with correctly overlaid text and mouse-controlled camera (centre right), then a much later iteration running in a custom physical enclosure (far right).}
  \Description{.}
  \label{img:tracker-opengl}
\end{figure}

This case study took place over a single, longer session of around 85 minutes. Unlike the other studies, I spent more iterations making finer adjustments to the user interface as I had quite specific ideas about how it should look and work. You can see in the timeline visualisation that there were 10 prompts in total. Figure \ref{img:tracker-opengl} shows the original tracker interface then two versions from the development of the new, 3D OpenGL interface, as well as a later iteration running in a custom hardware setup. In the first prompt, which took Codex around 10 minutes to respond to, I included references to examples in the libs/JUCE/examples folder which the agent could examine to find out how to add OpenGL capability to a JUCE program. I also attached a screenshot of the current, 2D interface. I did not provide much detail on how the keyboard controls worked but referred it to a particular C++ class in the project which contained the implementation, so unlike the other case studies, I was using specialised knowledge of my codebase in my prompts. 

The result of the first prompt was a non-working build so I prompted Codex to run the cmake command and fix any errors, which it did in about 30 seconds, resulting in the top right image in figure \ref{img:tracker-opengl}. At this stage, the 3D interface was in place, with the playhead cursor and edit cursor moving correctly but the data for the steps in the sequence was overlaid in 2D instead of being overlaid on the surfaces of the blocks. 
I sent Codex a screenshot of the interface as it was along with a new prompt and it was able to correctly overlay the text on the surface of the blocks. The implementation was interesting in that it generated a cached texture map of the text and overlaid it on the blocks. 

In the next phase I instructed the agent to add mouse control over the camera so I could zoom in and out and change perspective. In 30 seconds, it had completed this task successfully, resulting in the interface shown on the bottom left of figure \ref{img:tracker-opengl}. I returned later to the project for a few more interactions and was able to work towards the interface shown running in its custom hardware in the bottom right of figure \ref{img:tracker-opengl}. To get to this version, I used a combination of prompting and manual code editing. In some of the prompts, I asked Codex to explain how its code worked so I could find where to make my edits. I also instructed it to refactor the code to a more modular style. So, this case study stands apart from the others in that I did make use of my C++ skillset in various ways to speed things up. 
For example, using a natural language prompt to adjust a parameter that controls the decay rate of the orange glow in the boxes seemed much less efficient than asking where that parameter was and editing it myself. 

To summarise, this case study had similar patterns to the other two - starting with a solid foundation in the project folder, identifying examples that Codex could build on, the big prompt at the start followed by multiple smaller prompts. But I added some more techniques such as requesting explanation and manually editing code. 

\section{Discussion}

I started the paper by identifying two themes which would be of interest to the NIME community: the problem of longevity and interoperability, and the challenge of lowering the barrier to entry for the potentially large community of non-programmers who wish to experiment with the creation of new music technology. My aim was to explore the potential of agentic software engineering to address these challenges. 

In the Continuator case study, I was able to convert an SDMI written in Python and which depended on package installs and virtual environments, into a native plugin that would load in any DAW. I was able to do so only using natural language. In the Music Mouse case study, I was able to re-create an SDMI using only a user manual and some screenshots, again with natural language prompts. I did need some music technology domain knowledge, for example, around MIDI events and timing and generally understanding what plugins were and how they operated. I did not need any deep C++ knowledge.

Considering the theme of lowering the barrier for non-programmers, it is clear that it is possible to develop plugins from natural language, but you do need a well-designed starter program, and you do need to have a set of somewhat complex tools installed correctly beforehand. 

% But once those things are in place, my previous experience with the JUCE framework and my templates in educational contexts is that they are very portable, cross-platform and easy to install. So adding the ASE tools on top looks to be a desirable setup for non-programmers who wish to experiment with plugins. 

One problem for non-programmers is failing builds with obscure error messages, but Codex can deal with this, including explaining how it did so. More advanced, yet critical plugin concepts such as thread-safe data sharing between audio and GUI modules can be implemented through natural language, if beginners understood the basics of these ideas and how to request them. 

% Then looking ahead, to work towards stable, `production-ready' plugin software, it would be necessary to have an awareness of concepts such as thread safety, parameters and state saving. In other experiments I have found that mentioning a suitable design pattern for sharing data between threads (e.g. atomics or mutexes) or saving state (e.g. JUCE's AudioProcessorValueTreeState) is enough to achieve this, without needing to dive into the code. So for non-programmers, they would need to have some generic prompts they can use to get Codex to add these features. 

In the third case study, I showed how it is possible to work on a reasonably complex, existing audio program using ASE techniques. Here my familiarity with the codebase naturally led me to cut in and edit the code by hand sometimes, but the ASE tools assisted here as they could also explain to me how the code worked and indicate where I could make my edits. Using the tools to refactor also made the AI-generated code easier to understand. I was not very familiar with OpenGL programming, and I enjoyed learning some new techniques in OpenGL. So, for the intermediate or advanced programmer, or someone wishing to re-animate a dead SDMI's codebase there is plenty of value here too. 

\section{Conclusion and future work}

In this paper, I have presented three case studies wherein I used agentic software engineering techniques to re-create an existing SDMI from its user manual, translate an existing SDMI from Python to C++, and to develop a new user interface. I have considered how my findings relate to the problems of longevity and interoperability in SDMIs as well as the barrier to entry for non-programmers to SDMI creation. There are many areas to explore in ASE for audio software. I have chosen not to dive into ethical aspects here, but certainly others might find this a rich vein to explore. An interesting direction for my work will be developing workshop materials to support non-programmers in developing their own plugins. Another area with potentially high impact is the creation of a set of plugins that re-implement well known SDMIs from the literature, which would otherwise be lost in obscurity. 

\section{Acknowledgements}

I would like to acknowledge the fantastic work of Laurie Spiegel and Francois Pachet and team who developed the original Music Mouse and Continuator systems. Whilst Codex was used extensively to carry out the research, LLM technology was not used in any way to prepare the text in this paper, aside from assisting in generating the timeline visualisation. 

% \section{References}

\bibliographystyle{ACM-Reference-Format}
\bibliography{myk-ase-music-software-2026}

\end{document}